\documentclass[useAMS]{nature}
\usepackage{hyperref} 
\usepackage[sort,numbers]{natbib}
\usepackage{xcolor}
\usepackage[normalem]{ulem}
\usepackage{amsmath}
\usepackage{mathtools}  
\usepackage{graphicx}
\usepackage{caption}
\usepackage{bm}

\makeatletter
\let\saved@includegraphics\includegraphics
\AtBeginDocument{\let\includegraphics\saved@includegraphics}

\makeatother

\hypersetup{
    colorlinks=true,
    linkcolor=blue,
    filecolor=magenta,      
    urlcolor=cyan,
    citecolor=blue
}

\usepackage[percent]{overpic}
\usepackage{times}
\usepackage{mathabx}
\usepackage{subfigure}
\usepackage{natbib}
\usepackage{journals}
\usepackage{txfonts}
\usepackage[utf8]{inputenc}
\usepackage[overload]{empheq}
\usepackage{verbatim}

\usepackage{xcolor}

\renewcommand{\sout}[1]{\unskip}

\newcommand{\kms}{$\rm km\,s^{-1}$}

\title{A large, long-lived, slowly-expanding superbubble across the Perseus Arm}

\author
{Bingqiu Chen$^{1*}$, Guangxing Li$^{1*}$, Haibo Yuan$^{2,3}$, Maosheng Xiang$^{4,5}$, Jixuan Zhou$^{6}$, Pinjian Chen$^{4,5}$, Martin Krause$^7$, Ashley Coombs$^7$ \\
\small $^{1}${South-Western Institute for Astronomy Research, Yunnan University, Kunming, Yunnan 650091, P.\,R.\,China}\\
\small $^{2}${Institute for Frontiers in Astronomy and Astrophysics, Beijing Normal University, Beijing 102206, P.\,R.\,China}\\
\small $^{3}${School of Physics and Astronomy, Beijing Normal University, Beijing 100875, P.\,R.\,China}\\
\small $^4${National Astronomical Observatories, Chinese Academy of Sciences, Beijing 100012, P.\,R.\,China}\\
\small $^5${University of Chinese Academy of Sciences, Beijing 100049, P.\,R.\,China}\\
\small $^6${School of Physics and Astronomy, Cardiff University, Queen's Buildings, The Parade, Cardiff CF24 3AA, UK}\\
\small $^7${Centre for Astrophysics Research, School of Physics, Astronomy and Mathematics, University of Hertfordshire, College Lane, Hatfield, Hertfordshire AL10 9AB, UK} \\
\small $^*${Corresponding author. E-mails:bchen@ynu.edu.cn (BQC), gxli@ynu.edu.cn (GXL)}\\
}

\begin{document}

\maketitle


{\abstract{Stellar feedback is a crucial mechanism in galactic evolution, as demonstrated by the widespread bubbles observed with JWST. In this study, we combine data from Gaia and LAMOST to obtain a sample of young O-B2 stars with full three-dimensional velocity information. Focusing on the largest known superbubble in the Milky Way, we identify groups of O-B2 stars at its periphery, exhibiting a transverse velocity of 25.8\,\kms\ and an expansion velocity of 6.2\,\kms. Using these velocities, we calculate a crossing time $t_{\rm cross}\approx$ 20\,Myr and an expansion timescale $t_{\rm expansion }\approx$ 80\,Myr. We estimate a survival timescale $t_{\rm survival} \approx$ 250\,Myr and a supernova interval $t_{\rm SN} \approx$ 0.1\,Myr. Together with the Galactic shear timescale $t_{\rm shear} \approx$ 30\,Myr, these values satisfy $t_{\rm SN} < t_{\rm shear} < t_{\rm survival}$. The energy and momentum from supernovae are sufficient to sustain the bubble's growth against ambient pressure. This indicates that repeated supernovae replenish energy faster than shear and turbulent distort the cavity. Our analysis classifies the Giant Oval Cavity as a large, quasi-stationary superbubble, similar to the Phantom Bubble observed by JWST, stabilised by the interplay between stellar feedback and Galactic disk dynamics.}}



\maketitle
\section*{Introduction}

The structure and kinematics of spiral arms are fundamental to understanding the formation and evolution of the Milky Way \citep{Sellwood2002}. Spiral arms play a critical role in star formation by concentrating gas and dust into regions of high star forming activity \citep{Kalberla2009, Hou2009, Chen2019}. Investigating these structures provides key insights into the Galaxy's overall dynamics and the processes driving its evolution \citep{Block1999, Seigar2008, Binney2008, Kendall2015}. Both observational studies and theoretical models suggest that spiral arms are dynamic and transient features that significantly influence the spatial distribution and motion of stars and interstellar matter \citep{Sellwood2022, Li2022}.  

Star formation in the Milky Way begins with the collapse of molecular clouds, but this process is strongly regulated by stellar feedback \cite{Scannapieco2012,Stinson2013,Agertz2013}. Stellar feedback disrupts molecular clouds, thereby influencing the efficiency of star formation \citep{Williams1997, Luisi2021}. It also injects energy into the interstellar medium (ISM), sustaining turbulence and driving large-scale processes within the ISM \citep{Walch2012Dispersal, Pabst2019Disruption, Pabst2020Expanding}. The impact of stellar feedback is particularly evident in the widespread presence of bubbles observed throughout galactic disks, as recently highlighted by {James Webb Space Telescope (JWST)} observations \citep{Williams2024}. The size distribution of the bubbles detected by JWST exhibits a secondary peak and a trailing extension at large radii, with these features becoming more pronounced at greater galactocentric distances \citep{Watkins2023}. However, their physical properties remain poorly understood due to observational limitations. Superbubbles in the Milky Way, particularly those in the Solar neighborhood, thus provide rare, kinematically resolved benchmarks for studying superbubble evolution.

The Sun is located within a superbubble known as the Local Bubble, which is associated with the Local Arm of the Milky Way \citep{Chen2019b, Lallement2019, Pelgrims2020, Zucker2022}. By analyzing the three-dimensional (3D) distribution and motions of young stars, dust, and gas, we can identify large-scale structures such as superbubbles. These structures are shaped by the combined effects of stellar winds and supernova explosions, offering valuable insights into the feedback mechanisms that sculpt the ISM and influence the evolution of spiral arms. Despite their importance, comprehensive studies of superbubbles in spiral arms beyond the Local Arm remain limited. Recent advancements in observational astronomy, particularly through large-scale surveys such as {Large Sky Area Multi-Object Fiber Spectroscopic Telescope (LAMOST)} \citep{Cui2012, Luo2015} and Gaia \citep{Gaia2016, GaiaDR3}, have significantly enhanced the accuracy of stellar position and velocity measurements in the Milky Way. These improved datasets have deepened our understanding of spiral arm structures and underscored the role of young massive stars as key tracers of Galactic dynamics. In particular, the Perseus Arm—one of the Milky Way’s major spiral arms and the closest spiral arm in the outer Galaxy—has garnered attention due to its complex interactions and unique kinematic properties.

\section*{Results}

\subsection{O-B2 Stars in the Perseus Arm}

We have compiled a comprehensive catalog of 369 very young {(age $< 20\,\mathrm{Myr}$)}, massive O- and early B-type (O-B2) stars, using data from the LAMOST and Gaia surveys. These stars are associated with the Perseus arm, located within 0.5\,kpc of the arm's model in the Galactic $XY$ plane. The spatial distribution of these stars exhibits clear clustering, corresponding to the structure of the Perseus arm, as shown in Figure~\ref{xykine}. The 3D positions and velocities of these stars provide significant insights into the kinematic properties of the Perseus arm and the underlying mechanisms that influence the spatial distribution and motion of stars and interstellar matter.

In the outer Galaxy, the Perseus arm is primarily traced by six prominent OB associations, spanning the Cassiopeia, Perseus, Auriga, Gemini, and Orion regions. The average peculiar velocity of stars within the Perseus arm is approximately 16\,km\,s$^{-1}$ in the $XY$ plane, consistent with velocities observed in high-mass star-forming regions via maser measurements \cite{Reid2014}. These peculiar motions allow us to explore dynamic trends within each OB association. Our analysis reveals the presence of the largest known superbubble in the Milky Way, referred to as the Giant Oval Cavity \citep{vergely2022}. This structure spans across the Perseus arm, with a diameter exceeding $1\,\mathrm{kpc}$. It is primarily bordered by the Cassiopeia and Auriga OB associations in the Galactic plane, while the Perseus OB association lies below the plane. 

\subsection{The Giant Oval Cavity as a Slow-expanding Superbubble}

\begin{figure*}
\centering
\includegraphics[width=0.78\textwidth]{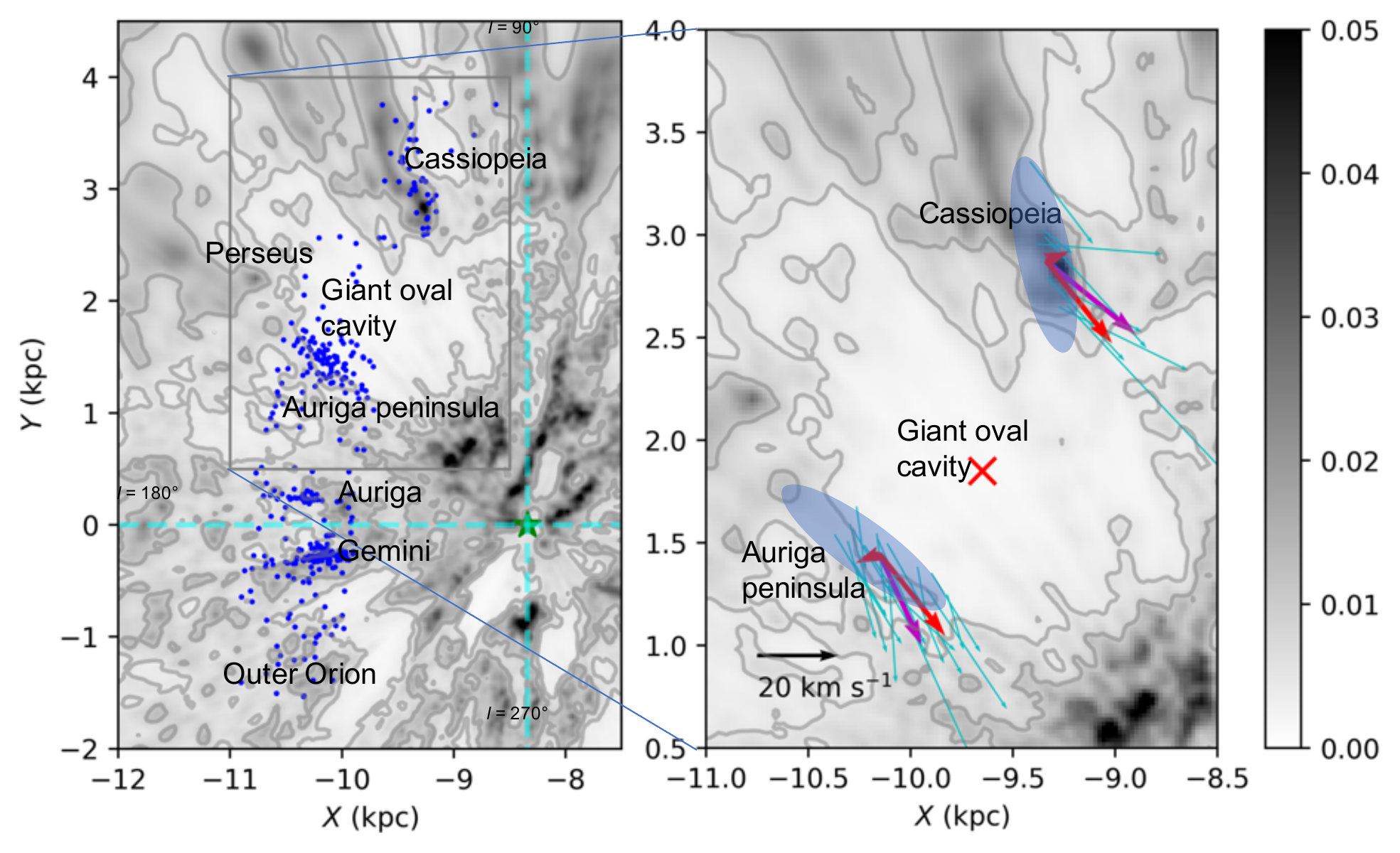}
\caption{ \textbf{Face-on view of O-B2 stars kinematics and gas distribution.}  
{Left panel}: {Spatial distribution of O-B2 stars (blue dots) within 0.5\,kpc of the Perseus Arm in the Galactic $X$$Y$ plane. The Sun (green pentagram) is positioned at $(X, Y) = (-8.34, 0)$\,kpc. OB associations follow the nomenclature of K. Jardine (\url{http://gruze.org/galaxymap/map_2020/}). Cyan dashed lines indicate Galactic longitude directions $l = 90^\circ$, $180^\circ$, and $270^\circ$.  }
{Right panel}: A zoomed-in view of the $X$$Y$ velocity vectors of O-B2 stars near the Galactic plane, focusing on the region surrounding the Giant Oval Cavity (cyan arrows). Median velocities of the Cassiopeia and Auriga Peninsula groups (magenta arrows) are decomposed into a common peculiar velocity and two oppositely directed expansion components (red arrows). Velocities are shown for stars within 100\,pc of the plane, excluding a few outliers. The black arrow (lower left) denotes a 20\,\kms\ reference velocity. Blue ellipses delineate the Cassiopeia and Auriga Peninsula groups. The grayscale map (ranging from 0-0.5\,mag\,pc$^{-1}$ ) and contours (0.005 and 0.01\,mag\,pc$^{-1}$) show differential extinction $\delta A_0$ at 550\,nm from \cite{vergely2022}. The cavity center, defined as the centroid of the low-dust region ($\delta A_0 < 0.005$\,mag\,pc$^{-1}$), is marked by a red cross. {Source data are provided as a Source Data file.}}
\label{xykine}
\end{figure*}

The Giant Oval Cavity extends across Galactic longitudes of approximately $100^{\degree}$ to $150^{\degree}$, as shown in Figure~\ref{xykine}. Within this region, at $|Z| < 100\,\mathrm{pc}$, the Cassiopeia and Auriga Peninsula groups exhibit coherent $XY$ motions directed outward from the cavity. This pattern strongly suggests an expansion phase for the superbubble. The observed stellar velocities result from a combination of their peculiar velocities and the expansion velocity of the surrounded cavity. To quantify these motions, we calculated the median velocities of the Cassiopeia and Auriga Peninsula groups and decomposed them into two components: a shared peculiar velocity and {two oppositely directed} components representing expansion from the cavity center. This decomposition is depicted in the right panel of Figure~\ref{xykine}. {From our analysis, we derived an expansion velocity of $v_{\rm expansion} \approx 6.2\pm2\,\mathrm{km\,s^{-1}}$ and a shared peculiar velocity of $v_{\rm parallel} \approx 25.6\pm3\,\mathrm{km\,s^{-1}}$.}  The expansion velocity of the Giant Oval Cavity is consistent with measurements for the Local Bubble, which has an expansion velocity of approximately $5.5\,\mathrm{km\,s^{-1}}$ \cite{Li2022}. However, the Giant Oval Cavity is significantly larger, with a diameter of $\sim1\,\mathrm{kpc}$ compared to the $\sim200\,\mathrm{pc}$ diameter of the Local Bubble \cite{Pelgrims2020}. 

\begin{figure*}
\centering
\includegraphics[width=0.79\textwidth]{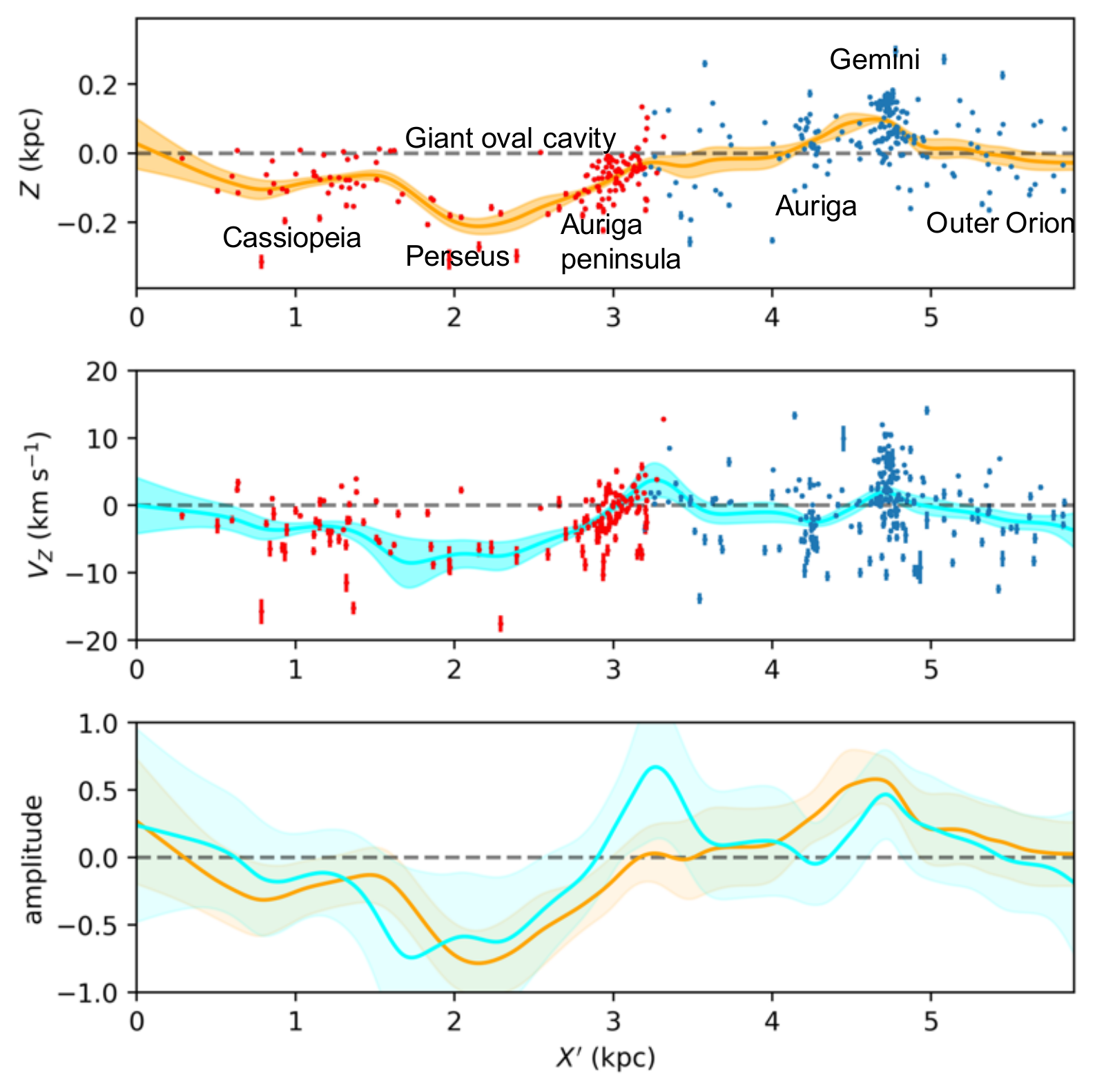}
\caption{{\bf Side view of the superbubble.} 
{The upper two panels display the vertical positions ($Z$) and vertical velocities ($V_Z$) along the Perseus arm. The $X'$ axis originates from ($X,~Y$) = ($-$8.8, 4.2)\,kpc and follows the orientation of the Perseus arm. O-B2 stars located within the Perseus arm and the vicinity of the superbubble are shown in blue and red, respectively. Error bars indicate the uncertainties in position and velocity measurements. The orange and cyan curves represent the smoothed profiles of $Z(X')$ and $V_Z(X')$, with shaded regions indicating the $1\sigma$ confidence intervals. For comparison, the bottom panel shows the normalized smoothed profiles. Grey dashed lines denote $Z = 0$ and $V_Z = 0$.}  {Source data are provided as a Source Data file.}}
\label{xpz}
\end{figure*}

In Figure~\ref{xpz}, we provide an edge-on view of the selected O-B2 stars. The Auriga peninsula, Perseus, and Cassiopeia OB groups form a continuous arc-like structure. The Auriga peninsula group is predominantly located near the Galactic plane, at $Z \approx 0$, while the Perseus group is situated below the plane, at $Z < 0$, forming the southern segment of the Giant Oval Cavity. The Cassiopeia group exhibits a vertical distribution rising to the plane. This arc is primarily found on the southern side of the Galactic plane, extending to approximately $Z \sim -270$\,pc at its southernmost point, which is significantly higher than the typical scale heights of the Milky Way's molecular gas disk, estimated to be 40--70\,pc \cite{Heyer2015, Guo2021, Li2024}. 

Additionally, Figure~\ref{xpz} shows the vertical velocity field associated with this large arc. Our analysis reveals a consistent vertical velocity pattern, with minimal oscillatory behavior. The O-B2 stars in the Auriga Peninsula group exhibit near-zero vertical velocities, while those in the Perseus group have notably negative velocities, indicating a rapid outward movement away from the Galactic plane. The Cassiopeia group shows a clear gradient in vertical velocities, with stars moving away from the plane at decreasing speeds as their distance from the Galactic plane decreases. A comparison of the vertical velocity and positional distributions reveals a striking agreement, with no discernible phase difference, suggesting that the arc is in a stable expansion phase. The estimated vertical expansion velocity of the superbubble is approximately 8.5\,km\,s$^{-1}$, which is comparable to its horizontal expansion velocity. 

The Giant Oval Cavity, identified as a $\sim$1\,kpc-scale region with reduced dust density \cite{vergely2022}, is likely the largest known superbubble in the Milky Way disk.  It represents an extreme example of stellar feedback-driven structures in disk galaxies. Interest in such large superbubbles has grown considerably following the identification of many similar features in external galaxies like NGC\,628 \cite{Watkins2023}. Several nearby stellar groups -- specifically the Cassiopeia, Perseus, and Auriga Peninsula -- exhibit positions and motions consistent with expansion driven by feedback from this cavity. In contrast, other groups along the Perseus arm, including Auriga, Gemini, and Outer Orion, show no evidence of such organized expansion. The absence of coherent radial motion in these groups, combined with the precise geometric alignment of expanding stars along the cavity boundary, suggests that these expansion patterns are not the result of random Galactic disk dynamics. Instead, they appear to be a direct consequence of the cavity’s influence. This disparity suggests a direct link between the ISM structure and the dynamics of associated stellar populations.

Observational data reveal a coherent expansion pattern, an organized velocity structure, and spatial undulations along the cavity’s boundary. These features support the interpretation of the cavity as a uniformly expanding superbubble. With an approximate diameter of 1\,kpc and a current expansion velocity of $\sim$6.2\,km\,s$^{-1}$, its estimated dynamical age, based on a simple $R / v$ calculation, is approximately $t_{\rm expansion} \approx \frac{500\,\mathrm{pc}}{6.2\,\mathrm{km\,s^{-1}}} \approx 80\,\mathrm{Myr}$. This makes it one of the oldest known superbubbles in the Galaxy, approximately four times older than the Local Bubble.

\subsection{Theoretical Framework for Bubble Stability}

We propose that the Giant Oval Cavity belongs to a class of quasi-stationary bubbles whose boundaries are sustained by a dynamical balance between external gas inflow and energy input from supernovae occurring within the cavity. The cavity is sufficiently large that numerous supernovae are expected to explode inside it before Galactic turbulence and shear can erase the structure. In galactic disks, molecular gas has a typical velocity dispersion of about {4}\,km\,s$^{-1}$ \cite{Marasco2017}. This velocity sets the rate at which turbulent gas flows can erode bubble boundaries. The corresponding survival timescale of a bubble is given by
\begin{equation}
    t_{\rm survival} = \frac{l_{\rm bubble}}{\sigma_{\rm v, disk}} \;,
\end{equation}
{where $l_{\rm bubble}$ is the characteristic size (i.e. diameter) of the bubble} and $\sigma_{\rm v, disk}$ is the velocity dispersion of the surrounding medium. Applying this to the Local Bubble and the Giant Oval Cavity yields:
\begin{equation}
    t_{\rm survival,\, Local\, Bubble} \bm{\approx} 50\,\mathrm{Myr} \;, 
\end{equation}
\begin{equation}
    t_{\rm survival,\, Giant\, Oval\, Cavity} \bm{\approx} 250\,\mathrm{Myr} \;.
\end{equation}
A bubble can retain its structure if a supernova explodes within it before it is eroded by turbulence. To estimate the supernova timescale, we assume a surface density of $5\,M_{\odot}\,\mathrm{pc}^{-2}$ \cite{Miville2017} and a gas depletion time of 2.5 Gyr \cite{Bigiel2011} for the solar neighborhood. This implies a star formation rate of $2\,M_{\odot}\,\mathrm{pc}^{-2}\,\mathrm{Gyr}^{-1}$. Assuming one supernova occurs per $150\,M_{\odot}$ of star formation \cite{Scannapieco2004}, the local supernova rate is:
\begin{equation}
    n_{\rm SN} = 0.01\,\mathrm{pc}^{-2}\,\mathrm{Gyr}^{-1} \;.
\end{equation}
The time between supernova events within a bubble of size $l_{\rm bubble}$ is then given by:
\begin{equation}
    t_{\rm SN} = \frac{1}{l_{\rm bubble}^2 n_{\rm SN}} \;.
\end{equation}
Substituting characteristic sizes gives
\begin{equation}
t_{\rm SN, ~Local ~Bubble} \bm{\approx} 2.5\,\mathrm{Myr} ,
\end{equation}
\begin{equation}
t_{\rm SN, ~Giant ~Oval ~Cavity} \bm{\approx} 0.1\,\mathrm{Myr} .
\end{equation}
For the Local Bubble, this agrees with the $^{60}$Fe isotope records of nearby supernovae $\sim$2--3\,Myr ago \cite{Opher2024} and with its ongoing slow expansion \cite{Zucker2022}. For the Giant Oval Cavity, the implied supernova loading ratio,
$\frac{t_{\rm survival}}{t_{\rm SN}} \bm{\approx} 2500$, 
indicates that a very large number of supernovae are expected to occur within the cavity over its survival time.

\begin{figure*}
    \centering
    \includegraphics[width=1\linewidth]{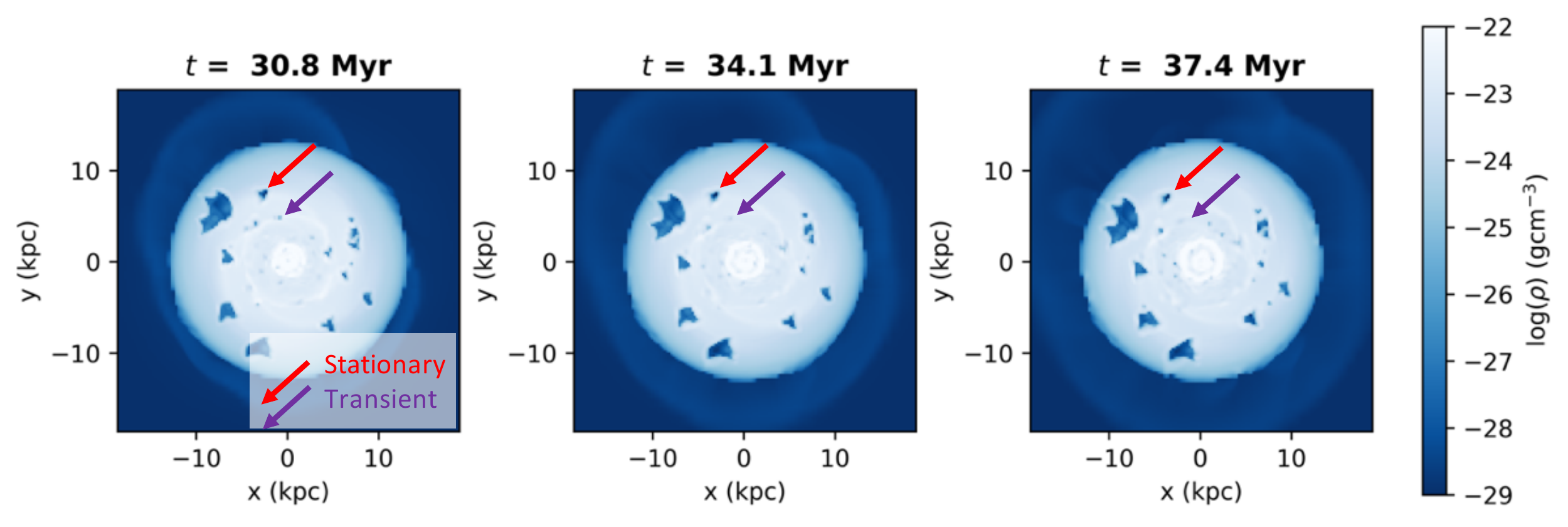}
    \caption{{\bf Evolution of superbubbles in galaxy-scale simulations.} Midplane gas density from simulations by \cite{Rodgers2019} and \cite{Krause2021} is shown. The results demonstrate that large, stationary bubbles {-- such as the one marked by the red arrow --} maintain their structure through a balance between gas inflow and local stellar feedback, while smaller bubbles {(purple arrow)} are quickly refilled with gas and dissipate unless a supernova occurs shortly after formation.  {Source data are provided as a Source Data file.}}
    \label{fig:simu}
\end{figure*}

This theoretical picture is supported by numerical simulations from \cite{Rodgers2019, Krause2021}, which model galaxies similar to the Milky Way. In these simulations, supernovae are introduced at rates consistent with the expected star formation activity of the interstellar medium. The results, shown in Figure~\ref{fig:simu}, indicate that small bubbles are typically short-lived, as they become filled with dense gas before the next supernova can occur. However, larger bubbles—often located in the outer regions of galaxies—remain stable over extended periods. 

{The stability of the Giant Oval Cavity involves two primary criteria: temporal and energetic. We compare the supernova replenishment time, $t_{\rm SN}$, with the timescale for differential rotation to distort the cavity, $t_{\rm shear} \approx 30\,\mathrm{Myr}$ \cite{Zhou2022}, and with the turbulent erosion time, $t_{\rm survival}$. Supernovae must recur on a timescale shorter than both competing processes. Our calculated values, $t_{\rm SN} < t_{\rm shear} < t_{\rm survival}$, meet this condition, indicating that feedback events are sufficiently frequent to prevent the bubble's disruption. The combined energy and momentum input from supernovae must be sufficient to create the cavity by overcoming the gravitational binding energy and turbulent pressure of the ambient gas, and to offset radiative losses to maintain the cavity’s expansion over time. As detailed in the Methods, the Giant Oval Cavity requires approximately 400 supernovae to be powered. Our estimated supernova loading of the system satisfies this requirement, confirming that the energy and momentum budgets are ample. Meeting both the temporal and energetic criteria indicates that the Giant Oval Cavity is a long-lived, feedback-driven structure undergoing quasi-stationary expansion. In this regime,} Galactic shear deforms the cavity significantly without erasing it completely, while the bubble evolves through concurrent expansion, contraction, and distortion. Its boundary maintains a dynamical equilibrium between shear forces, supernova feedback, and turbulent gas inflow. This explains the irregular boundaries seen in simulations ({Figure}~\ref{fig:simu}).

Several additional factors contribute to this picture. The bubble's low-density interior likely reduces star formation rates compared to Galactic averages, potentially decreasing internal supernova rates. However, triggered star formation at cavity edges counteracts this by enhancing stellar birth in compressed shells, which would further shorten $t_{\rm SN}$. This aligns with observations showing elevated star formation rates in the Giant Oval Cavity region \cite{Elia2022, Zari2023}, consistent with feedback-triggered star formation scenario.

Moreover, bubbles exceeding the Milky Way's gas scale height ($\sim$100 pc) develop chimneys that vent hot gas vertically. Although $t_{\rm SN} \ll t_{\rm shear}$ implies feedback-dominated expansion, radial growth might halt when cavities surpass this scale height. For the Giant Oval Cavity, vertical venting converts radial momentum into vertical outflows, limiting lateral expansion while maintaining structural coherence through continuous feedback. Heat conduction \cite{El2019} also contributes to energy loss, but for larger structures, the relevant surface area—approximately $r_{\rm bubble} h_{\rm disk}$—is proportionally smaller relative to their volume, reducing the overall impact.

Observations of young OB stars located beneath the cavity reveal vertical velocities directed away from the Galactic plane, consistent with momentum injection from recent supernovae. This supports a feedback cycle in which massive star explosions initially clear surrounding gas to create the cavity, while subsequent shockwaves compress nearby clouds, triggering new star formation. Later generations of supernovae from these young stars then replenish the system with energy, sustaining its expansion. This process aligns with our discovery of the $\sim$10\,Myr-old clusters observed along cavity edges, whose velocities match feedback-driven gas motions.  We have conducted a detailed calculation of the number of supernovae required to form and maintain the Giant Oval Cavity. Accounting for the combined effects of energy injection, radiative loss, and ISM interaction, we estimate that approximately 400 supernovae are necessary. This estimate aligns with predictions from the Kennicutt--Schmidt relation \citep{Kennicutt1998} for star formation on kiloparsec scales. A complementary estimate, scaling the Local Bubble’s $\sim$ 20 supernovae \cite{Berghofer2002,Zucker2022} by the cavity’s fivefold larger radius, yields about 500 supernovae -- in close agreement with our detailed calculation. These feedback processes inject significant momentum into the ISM, driving large-scale structures such as gas chimneys that redistribute material vertically and may play a key role in shaping the interface between the Milky Way’s disk and halo.

In summary, large bubbles like the Giant Oval Cavity reach a quasi-stationary state where their boundaries are shaped by a dynamical balance between mass inflow, galactic shear, and internal supernovae. Shear continuously deforms the structure, while supernovae within the cavity counteract this by pushing material outward, maintaining a quasi-stationary boundary. These findings match galaxy-scale simulations showing that large cavities become long-lived, dynamically-evolving structures capable of migrating within host galaxies. This framework provides a clear explanation for the nature of giant cavities such as the Phantom Bubble recently observed by JWST.

\section*{Discussions}

\subsection{Ruling Out Alternative Explanations}

Alternative explanations for the origin of the cavity include the possibility that it is a purely gravitational structure, such as an unusually empty inter-arm region, or a formation related to Parker instability. In either case, the OB stars would have formed within a pre-existing underdensity, making it easier to clear material and potentially leading to a more recent creation of the cavity. However, the spatial correlation between OB stars and the cavity walls alone does not necessarily imply that the stars formed the entire structure.
If the cavity were simply a quiescent inter-arm region, its kinematics would be expected to follow the patterns of Galactic rotation and shear. However, the Cassiopeia and Auriga Peninsula groups exhibit radially outward velocities ($v_{\text{expansion}} \approx 6.2\,\mathrm{km\,s^{-1}}$), which deviate significantly from the expected motions in a passive shear environment. Additionally, the dust column density within the cavity ($\delta A_0 < 0.005\,\mathrm{mag\,kpc^{-1}}$) is an order of magnitude lower than that typically observed in inter-arm regions ($\sim 0.2\,\mathrm{mag\,kpc^{-1}}$; \cite{vergely2022}), suggesting an initial underdensity that would be unusually extreme. The associated stellar population also shows a distinct difference: stars near the cavity walls are young (ages $\leq 20\,\mathrm{Myr}$), whereas inter-arm regions are generally populated by older stars. These contrasts in both kinematics and stellar age strongly argue against a passive, pre-existing underdensity as the cavity's origin.

Considering the Parker instability scenario, theoretical models predict the formation of gas-dominated structures, characterized by  undular modes (wavelength $ \sim 1\,\mathrm{kpc}$; \cite{Parker1979,Rodrigues2016}). However, the Giant Oval Cavity displays a smooth, elliptical morphology that does not match these theoretical predictions. Moreover, the cavity is defined by significant dust evacuation and coherent vertical velocity gradients—features not typically produced by Parker instability, which primarily organizes gas without clear signatures of stellar feedback.

Further supporting this conclusion, most O-B2 stars in the Cassiopeia and Auriga Peninsula groups exhibit radially outward motions relative to the cavity center. Such coherent kinematics are unlikely to result from random star formation in a pre-existing void, where more isotropic velocity distributions would be expected. In contrast, other nearby OB groups (e.g., Auriga, Gemini, and Outer Orion) do not display similar velocity alignment. In summary, the combination of extremely low dust density, the presence of a young stellar population, the deviation from shear-driven kinematics, and the organized feedback-driven morphology collectively argue against the scenarios of a pre-existing underdensity or the Parker instability. The spatial and kinematic correlation between the OB stars and the cavity walls provides strong evidence that stellar feedback is the primary driver of the cavity's formation and its ongoing evolution.

\subsection{Feedback Cycle and Star Formation Implications}

Finally, the remarkably consistent $XY$ velocities of the Cassiopeia and Auriga Peninsula groups support the hypothesis that these groups originated from a single large molecular filament on the kpc scale \cite{Li2017, Li2022}. As this filament traversed the Giant Oval Cavity, it was acted upon by stellar winds and supernova feedback, which disrupted the filament and triggered star formation. The filament's two ends were pushed apart, eventually fragmenting into the Cassiopeia and Auriga Peninsula groups. This scenario explains both the observed velocity structure and the spatial distribution of these groups within the cavity. The stellar winds and supernova feedback not only disrupted the filament but also injected energy into the ISM, shaping the dynamics of the surrounding region. The fragmentation process highlights the role of feedback in dynamically influencing the formation and dispersal of star-forming regions on large scales.

\section*{Methods}\label{sec11}

\subsection{Data and Sample Selection}\label{sec11a}

This study utilizes data from LAMOST DR9 v1.0 \cite{Luo2015}, which includes both low- and medium-resolution spectra obtained by LAMOST between October 2011 and June 2021. For this work, we focus on the low-resolution catalogue, which contains over 10 million spectra. The low-resolution spectra have a resolution of $R \sim 1800$ and cover the optical wavelength range from 3800 to 9000\,\AA. Additionally, we incorporate data from Gaia DR3 \cite{GaiaDR3}, which provides astrometric solutions and broad-band photometric measurements ($G$, $G_{\rm BP}$, $G_{\rm RP}$) for approximately 1.5 billion sources.

We selected O- and early B-type stars from the combined LAMOST and Gaia datasets. First, we cross-matched these datasets using a 1.5\arcsec\ search radius. We then filtered stars with parallax uncertainties smaller than 20\%. Distances were calculated from Gaia DR3 parallaxes using a Bayesian approach \cite{Maz2005, Bailer2018}, with the posterior probability given by:
\begin{equation}
   p(d|\varpi) = d^2 \exp \left( -\dfrac{1}{2\sigma^2_{\varpi}} \left( \varpi - \varpi_{\rm zp} - \dfrac{1}{d} \right) \right) p(d),
\end{equation}
where $\sigma_{\varpi}$ and $\varpi_{\rm zp}$ are the uncertainties and global zero-point offset of Gaia parallaxes, respectively, and $p(d)$ is the prior on the space density distribution of the stars. We adopt a zero point of $\varpi_{\rm zp} = -0.026$\,mas from \cite{Huang2021}, and the Galactic structure model from \cite{Chen2017} as the prior.

\begin{figure*}
\centering
\includegraphics[width=\textwidth]{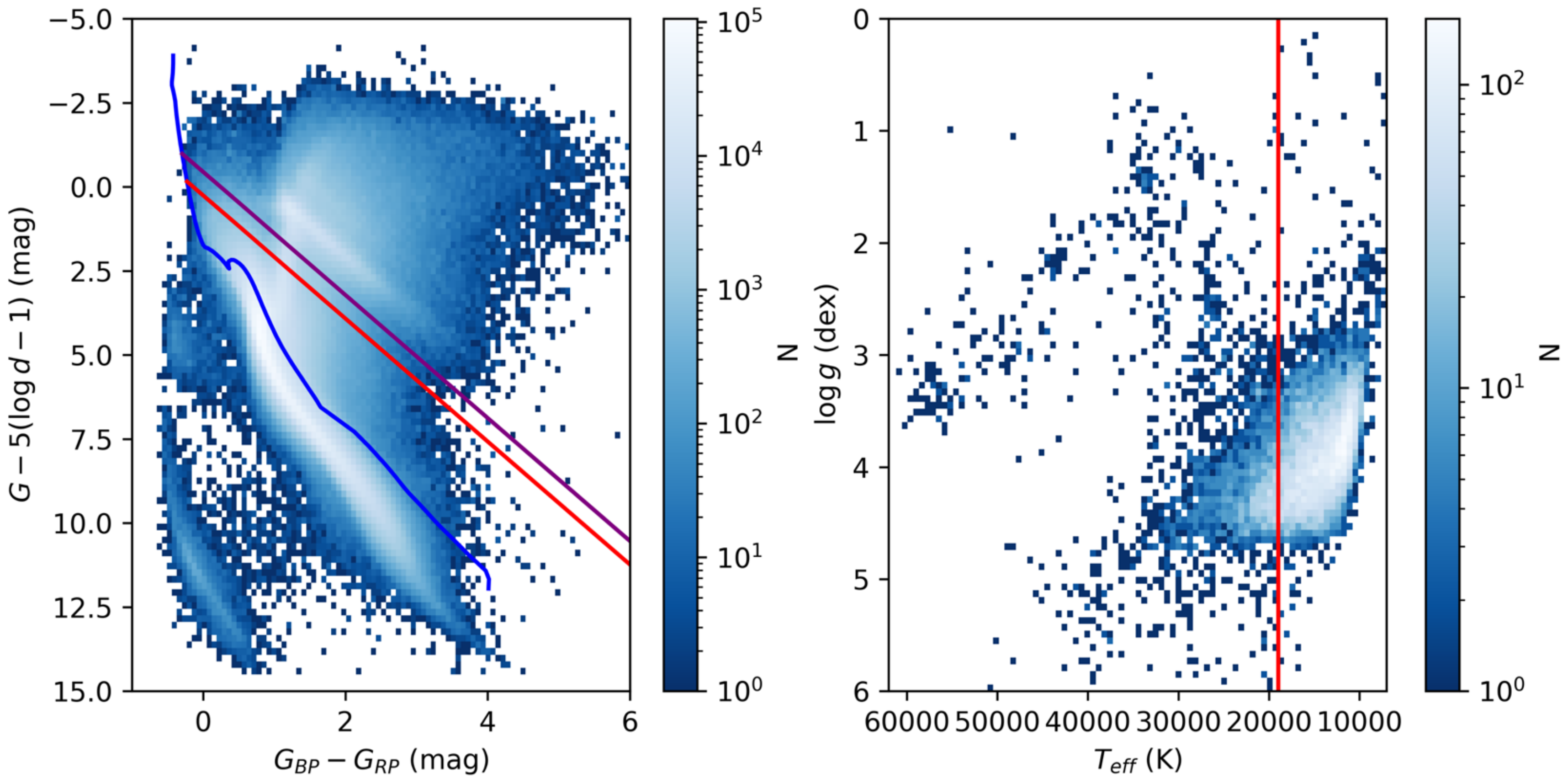}
\caption{\textbf{Selection of O-B2 stars.} {Left panel}: Gaia color versus `absolute' magnitude diagram of all LAMOST and Gaia stars with distance uncertainties smaller than 20\%. The blue curve represents the PARSEC isochrones \cite{Marigo2017} for main-sequence stars with an age of 10\,Myr. The red and purple lines correspond to the reddening curves for B5V and B3V stars, respectively. {Right panel}: Effective temperature versus surface gravity diagram of the selected OB stars. The vertical line marks $T_{\rm eff} > 19,000$\,K.  {Source data are provided as a Source Data file.}}
\label{obsel}
\end{figure*}

In the initial phase of our analysis, we excluded late-type main-sequence stars using the Gaia color-`absolute' magnitude diagram. The term `absolute' magnitude here refers to a magnitude that accounts for the distance modulus but does not include corrections for interstellar extinction. It is calculated as $G - 5 \log\,d + 5$, where $d$ denotes the distance. As shown in the left panel of Figure~\ref{obsel}, we systematically removed late-type stars based on their observed colors and `absolute' magnitudes. We retained only stars located above the B5V extinction curve, refining our sample to 57,375 stars. 

A significant portion of the sample consisted of late-type giants. The lack of extinction corrections led to some overlap between early B-type main-sequence stars and redder giants. To further refine the selection, we calculated spectral line indices from the LAMOST spectra for these stars. {We computed indices such as CaII K, H$\delta$, CN, Ca4227, G4300, H$\gamma$, Fe4383, He4388, Fe4531, H$\beta$, Ca4455, He4471, Fe4531, He4542, Fe4668, H$\beta$, Fe5015, Mg1, Mg2, Mg b, Fe5270, Fe5335, Fe5406, and Fe5709  \cite{Liu2015, Liu2019}. To derive spectral line indices, we first identify the line of interest and select two continuum regions on either side of it. A linear fit is then applied to the flux values in these regions to establish the pseudo-continuum level. Finally, the line index (i.e. equivalent width) is computed by integrating the flux difference between the observed spectrum and the pseudo-continuum across the line's wavelength range.} Using machine learning techniques, specifically random forest classfier, we identified OB stars among the candidates. Cross-referencing with the Simbad database, we selected 7,873 stars with identified spectral types, including 3,529 OB-type stars and 4,344 stars of other types. We then trained a random forest classifier with spectral indices as inputs and spectral types as outputs. The distinct spectral line features of OB stars allowed the classifier to achieve an accuracy of 97\%. Applying this classifier to the remaining candidates, we identified a final sample of 19,600 OB-type stars.

The primary objective of our study is to isolate very young {(with ages $t <$ 20\,Myr)} O- and early B-type stars (O-B2) to trace the superbubble structure of the Milky Way. For this purpose, we calculated atmospheric parameters for the 19,600 OB stars identified through spectral line indices using the HOTPAYNE pipeline \cite{Xiang2022}. As shown in the right panel of Figure~\ref{obsel}, the majority of the selected OB stars are main-sequence stars with temperatures exceeding 10,000 K. We specifically focused on stars earlier than B3 in spectral type, with effective temperatures $T_{\rm eff} > 19,000$\,K. This rigorous selection process resulted in a final sample of 1,608 unique O-B2 stars.

\subsection{Three-Dimensional Position and Velocity}\label{sec11b}

To determine the radial velocities of the refined O-B2 star sample, we analyzed their spectra obtained from the LAMOST telescope. Using theoretical template spectra for OB stars provided by TLUSTY \cite{TLUSTY}, we performed spectral fitting in the blue wavelength range (4000–5400\,\AA). This allowed us to accurately derive the line-of-sight velocities for each OB star. {To estimate the uncertainty in radial velocities, we used a {Monte Carlo (MC)} method based on the spectral uncertainties reported by LAMOST. Specifically, we added normally distributed noise to the flux values based on their respective errors, recalculated the radial velocities 100 times, and derived the standard deviation of the results as the radial velocity uncertainty.}

By combining the distances calculated from Gaia DR3 parallaxes, the radial velocities obtained from LAMOST, and the positional and proper motion data from Gaia DR3, we derived the three-dimensional positions ($X$, $Y$, $Z$) and velocities ($V_x$, $V_y$, $V_z$) for every star in the sample. In this coordinate system, the $X$-axis points towards the Galactic center, the $Y$-axis aligns with the direction of Galactic rotation, and the $Z$-axis points towards the North Galactic Pole. Similarly, the $U$-axis is directed towards the Galactic center, the $V$-axis follows the Galactic rotation, and the $W$-axis extends towards the North Galactic Pole. For this study, we adopted the following reference values for the solar neighborhood: the Sun's position is $(X_{\odot}, Y_{\odot}, Z_{\odot}) = (-8.34$, 0, 0.025)\,kpc \cite{Reid2014,Juric2008}, with local standard of rest (LSR) velocities of $(U_{\odot}, V_{\odot}, W_{\odot}) = (11.69, 10.16, 7.67)$\,\kms, Oort constants ($A$, $B$) = (16.31, $-11.99$)\,\kms\,kpc$^{-1}$, and the Sun's rotational velocity $V_{\rm LSR} = 231$\,\kms\ \cite{Wang2021}. Using these values, we calculated the peculiar velocities for each star by subtracting the LSR motion, Galactic rotation, and shear effects \cite{Li2022}.

The spatial distribution of the selected O-B2 stars in the Galactic $X$-$Y$ plane is shown in the left panel of  Figure~\ref{xydist}. Due to the observational footprint of the LAMOST survey, most of these stars are situated in the outer Galactic disk. The majority of the young O-B2 stars exhibit a clustered distribution, while those at larger distances appear more dispersed, likely due to greater uncertainties in the distance estimates. Among OB associations with well-determined distances within 3\,kpc, the spatial arrangement of stars clearly reflects the spiral arm structure \cite{Chen2019b}. In particular, we observe distinct concentrations of OB stars aligned with the Orion and Perseus arms of the Milky Way.

\subsection{O-B2 Stars Associated with the Perseus Arm}\label{sec11c}

We identified a subset of O- and early B-type stars associated with the Perseus arm of the Milky Way. This selection was based on the Perseus arm model described by \cite{Chen2019b}. Specifically, we selected stars located within 0.5\,kpc of the Perseus arm in the Galactic $X$-$Y$ plane. This criterion resulted in a sample of 369 O-B2 stars.

\begin{figure*}
\centering
\includegraphics[width=\textwidth]{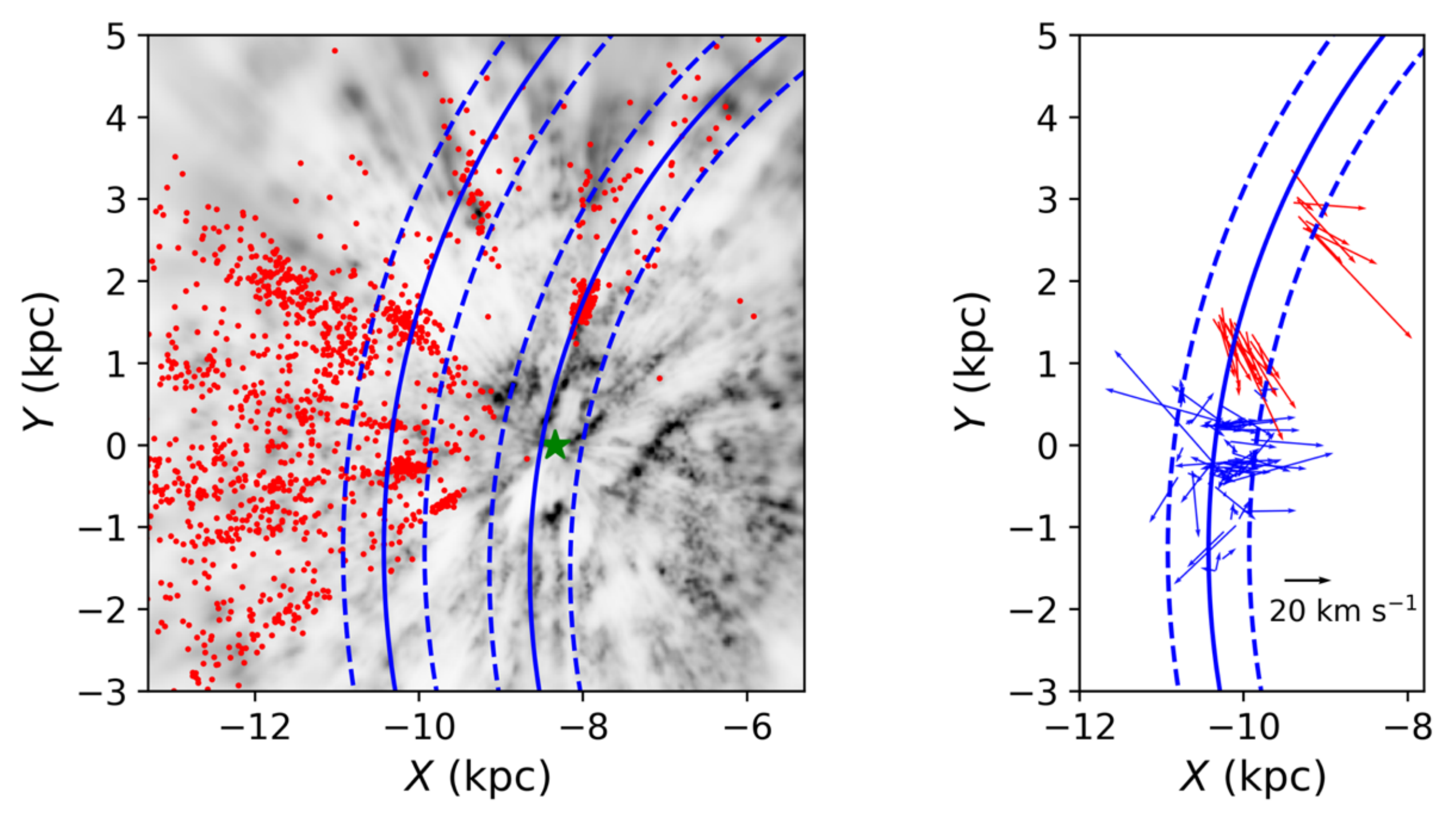}
\caption{\textbf{Distribution of O-B2 stars in the $XY$ plane.} {Left panel}: Spatial distribution of the selected O-B2 stars in the Galactic $XY$ plane. The Sun, located at $(X, Y)$ = ($-$8.34, 0)\,kpc, is marked by a green pentagram. The background grayscale represents the dust distribution from \cite{vergely2022}. Solid blue lines delineate the best-fitting spiral arm models for the Local and Perseus Arms from \cite{Chen2019b}, with dashed lines showing the 0.5\,kpc range around the arm models.  {Right panel}: Peculiar velocity vectors of O–B2 stars associated with the Perseus Arm and located within 100\,pc of the Galactic midplane. Stars near the Giant Oval Cavity are shown as red arrows, while other stars are shown as blue arrows. Some stars with exceptionally high velocities are not shown for clarity. The black arrow at the lower right represents a reference velocity of 20\,\kms.  {Source data are provided as a Source Data file.}}
\label{xydist}
\end{figure*}

The spatial distribution of these stars, as shown in the left panel of Figure~\ref{xydist}, supports their association with the Perseus arm. A detailed catalog of these stars is available online at the following URL: \url{https://paperdata.china-vo.org/diskec/obsamp/ob2perseus.fits}. This catalog offers comprehensive data for each star, including celestial coordinates, the signal-to-noise ratio of the LAMOST spectra, distance estimates, parallax values, proper motions, radial velocities, and associated uncertainties. 

The right panel of Figure~\ref{xydist} displays the peculiar velocity vectors of stars associated with the Perseus arm and located within 100\,pc of the Galactic midplane, in the $X$–$Y$ plane. These velocity patterns provide insight into the local kinematics of OB associations along the arm. A clear contrast emerges: stars near the Giant Oval Cavity—particularly those in the Cassiopeia and Auriga Peninsula regions (marked with red arrows)—exhibit coherent radial expansion away from the cavity center. In comparison, stars in other parts of the Perseus arm, such as Auriga, Gemini, and Outer Orion (blue arrows), display more irregular motions without any systematic radial expansion. 

\subsection{{Uncertainties of the Expansion and Peculiar Velocities}}

{To estimate the uncertainties of the expansion and peculiar velocities of the OB stellar groups associated with the Giant Oval Cavity, we employed a MC approach based on the uncertainties of the three-dimensional positions and velocities of the individual O-B2 stars. In this method, we perturbed the measured positions and velocities by adding random noise consistent with their observational errors and recalculated the expansion and peculiar velocities across 300 realizations. The standard deviations of these resultant velocities were adopted as the uncertainty estimates: approximately 2~km~s$^{-1}$ for the expansion velocity and 3~km~s$^{-1}$ for the peculiar velocity.}

\subsection{{Smoothing of Vertical Position and Velocity Profiles}}

{To derive smoothed profiles of the vertical positions ($Z$) and vertical velocities ($V_Z$) as functions of $X'$, we utilized the \texttt{smoothfit} Python package, which applies constrained optimization techniques for smooth curve fitting. Specifically, we performed one-dimensional smoothing spline fitting using \texttt{smoothfit.fit1d}. The fitting was conducted over the relevant range using 1,000 evenly spaced evaluation points, linear basis functions (degree = 1) to ensure piecewise continuity, and a regularization parameter $\lambda = 0.1$ to balance between fitting accuracy and smoothness.

To assess the uncertainties of the smoothed profiles, we implemented 1,000 nonparametric bootstrap resamplings of the original dataset. Each resampled dataset was smoothed using the same method, and the 1$\sigma$ confidence intervals were computed from the pointwise standard deviations of the ensemble of smoothed curves.}

\subsection{{Number of supernova Events}}\label{sec11e}

{To estimate the number of supernovae needed to create and maintain a superbubble in the Galactic disk, we modeled the bubble as an oblate spheroid with an in-plane radius of $R \bm{\approx} 500 \, \text{pc}$ and a vertical height of $h \bm{\approx} 100 \, \text{pc}$, reflecting the disk's thin structure. The volume $V$ of the bubble is calculated as $V = \frac{4}{3} \pi R^2 h$, yielding $V \bm{\approx}  3.1 \times 10^{63} \, \text{cm}^3$. Our energy budget includes three components: gravitational binding energy ($E_{\text{bind}}$), turbulent pressure energy ($E_{\text{turb}}$), and radiative cooling losses ($E_{\text{loss}}$). First, $E_{\text{bind}}$ is derived by the standard relation $E_{\text{bind}} = \frac{3}{5}\frac{G(\Sigma_g \pi R^2)^2}{R}$. Assuming a gas surface density of $\Sigma_g = 10 \, M_\odot \, \text{pc}^{-2} $, resulting in $E_{\text{bind}} \bm{\approx} 6.3 \times 10^{51} \, \text{erg}$. Second, $E_{\text{turb}}$ is computed as $E_{\text{turb}} = \mathcal{E}_{\text{turb}} V$. We have $\mathcal{E}_{\text{turb}} = \frac{1}{2} \rho_0 \delta v^2$, with  number density $n_0 = 1 \, \text{cm}^{-3}$ and velocity dispersion $\delta v = 10 \, \text{km s}^{-1}$, giving $E_{\text{turb}} \bm{\approx}  2.5 \times 10^{51} \, \text{erg}$. Third, $E_{\text{loss}}$ is calculated as $E_{\text{loss}} = n_e^2 \Lambda(T) V t_{\text{exp}}$, using electron density $n_e = 0.01 \, \text{cm}^{-3}$ (for dilute hot gas at $T \sim 10^6 \, \text{K}$), cooling function $\Lambda(T) = 10^{-22} \, \text{erg cm}^3 \, \text{s}^{-1}$, and expansion timescale $t_{\text{exp}} = 80 \, \text{Myr}$, yielding $E_{\text{loss}} \bm{\approx} 7.8 \times 10^{52} \, \text{erg}$. The total energy requirement is $E_{\text{total}} \bm{\approx} 8.7 \times 10^{52} \, \text{erg}$. Assuming each supernova releases $E_{\text{SN}} \bm{\approx} 10^{51} \, \text{erg}$ with a coupling efficiency of 20\%, this leads to a required number of supernovae as $N_{\text{SN}} \bm{\approx} 435$. 

This supernovae number aligns with the expected output from a stellar population with a mass of $M_* \sim 10^5 \, M_\odot$, based on a standard Salpeter initial mass function (IMF) that assumes an average of one supernova per $150 \, M_\odot$ of formed stars \cite{Scannapieco2004}. The Kennicutt-Schmidt (KS) relation \cite{Kennicutt1998} estimates a star formation rate surface density of: $\Sigma_{\text{SFR}} \approx 2.5 \times 10^{-4} \left( \frac{\Sigma_{\rm g}}{1\,M_\odot \, \text{pc}^{-2}} \right)^{1.4} \, M_\odot \, \text{yr}^{-1} \, \text{kpc}^{-2}$. For $\Sigma_g = 10 \, M_\odot \, \text{pc}^{-2}$ over a period of 80\,Myr, this results in a stellar mass of $M_* \sim 4 \times 10^5 \, M_\odot$, corresponding to approximately 2600 supernovae. This value is slightly higher but remains consistent with our estimate, considering uncertainties in supernova feedback efficiency and the stochastic nature of IMF sampling.}


{\noindent \small {\bf Data Availability} The Gaia DR3 is publicly accessible through the Gaia Archive. Spectroscopic data from the LAMOST DR9 are available on the official LAMOST data portal. A catalog containing detailed information of 369 O-B2 type stars, used in this study, can be accessed at the following URL: \url{https://paperdata.china-vo.org/diskec/obsamp/ob2perseus.fits}. {Source data are provided with this paper.}}

{\noindent \small {\bf Code Availability} The analysis in this work was carried out using several open-source Python packages, including \texttt{Astropy}, \texttt{smoothfit}, \texttt{emcee}, \texttt{galpy}, \texttt{scipy}, and \texttt{scikit-learn}. The visualization and data exploration tool \texttt{TOPCAT} was also employed.}

\section*{References}
\bibliographystyle{sn-nature}
\bibliography{perseusbubble}

{\noindent \small {\bf Acknowledgements} This work is partially supported by the National Natural Science Foundation of China 12173034 {(BQC)} and 12322304 {(BQC)}. We acknowledge the science research grants from the China Manned Space Project with no. CMS-CSST-2021-A09 {(BQC, HBY)}, CMS-CSST-2021-A08 {(BQC, HBY)}, and CMS-CSST-2021-B03 {(BQC, GXL)}. 
Guoshoujing Telescope (the Large Sky Area Multi-Object Fiber Spectroscopic Telescope LAMOST) is a National Major Scientific Project built by the Chinese Academy of Sciences. Funding for the project has been provided by the National Development and Reform Commission. LAMOST is operated and managed by the National Astronomical Observatories, Chinese Academy of Sciences.
This work presents results from the European Space Agency (ESA) space mission Gaia. Gaia data are being processed by the Gaia Data Processing and Analysis Consortium (DPAC). Funding for the DPAC is provided by national institutions, in particular the institutions participating in the Gaia MultiLateral Agreement (MLA). The Gaia mission website is https://www.cosmos.esa.int/gaia. The Gaia archive website is https://archives.esac.esa.int/gaia.}

{\noindent \small{\bf Author contributions}
B. Chen developed the initial concept,  conducted the analysis using LAMOST and Gaia data. B. Chen and G. Li led interpretation of the observational results, aided by H. Yuan. M. Xiang, J. Zhou and P. Chen focused on the OB stellar atmospheric parameters and velocity determinations. M. Krause and A. Coombs performed hydrodynamic simulations and co-developed the evolutionary model. All authors contributed to manuscript writing and revisions. }

{\noindent \small{\bf Competing Interests}
 The authors declare no competing interests.}

\section*{Figures}

\end{document}